# Moiré structures in twisted bilayer graphene studied by transmission electron microscopy

Tatiana Latychevskaia*, Conrad Escher, Hans-Werner Fink

Physics Department, University of Zurich,

Winterthurerstrasse 190, 8057 Zurich, Switzerland

*Corresponding author: tatiana@physik.uzh.ch

**ABSTRACT**

We investigate imaging of moiré structures in free-standing twisted bilayer graphene (TBG) carried out by transmission electron microscopy (TEM) in diffraction and in-line Gabor holography modes. Electron diffraction patterns of TBG acquired at typical TEM electron energies of 80 – 300 keV exhibit the diffraction peaks caused by diffraction on individual layers. However, diffraction peaks at the scattering angles related to the periodicity of the moiré structure have not been observed in such diffraction patterns. We show that diffraction on moiré structure can create intense diffraction peaks if the energy of the probing electrons is very low, in the range of a few tens of eV. Experimental diffraction patterns of TBG acquired with low-energy electrons of 236 eV exhibiting peaks attributed to the moiré structure periodicity are shown. In holography mode, the intensity of the wave transmitted through the sample and measured in the far-field can be enhanced or decreased depending on the atomic arrangement, as for example AA or AB stacking. Thus, a decrease of intensity in the far-field must not necessarily be associated with some absorption inside the sample but can simply be a result of a particular atomic arrangement. We believe that our findings can be important for exploiting graphene as a support in electron imaging.

# 1. INTRODUCTION

Moiré structure formed by superposition of periodical structures is highly sensitive to the local aperiodicities in the periodical structures. Electron transmission imaging of moiré structures created by superposition of single crystal metal films of a few hundred Angstroms thickness has been used to visualize and characterize dislocations since the early years of electron microscopy [1-3]. Recently, moiré structures created in twisted bilayer graphene (TBG) [4, 5] have become of a particulate interest for their anomalous electronic properties [6]. In 2011 Bistritzer and MacDonald predicted that under certain twisted "magic" angles, TBG should exhibit superconductivity [7], which was successfully measured at the first "magic" twist angle of 1.1° by Cao et al in 2018 [8].

Typically, moiré structures in TBG are visualized by scanning tunnelling microscopy (STM) which employs the close proximity between the conductive tip and sample [9, 10]. In the present study we investigate electron transmission microscopy applied for visualization of few layer twisted graphene where the transmitted electron wave is detected in the far field.

# 2. METHODS

## 2.1 Formation of moiré peaks in diffraction pattern of twisted lattices

A moiré pattern is formed when two periodical structures are superimposed on top of each other. The formed pattern is a periodical structure by itself, we will call it moiré structure. Diffraction on superposition of two periodical structures exhibit peaks caused by diffraction on each individual periodical structure but does not necessarily peaks caused by diffraction on the moiré structure. The reason is as follows. We consider a model where two periodical structures (lattices) are made up of identical absorbing particles. First, we assume that each particle has an infinitesimal size, and thus can be described by a $\delta$-function. It means that the transmission function of each lattice can be written as

$$t_i(x,y) = 1 - w_i(x,y), \quad (1)$$

where $w_i(x,y)$ is the two-dimensional Dirac comb-function, $i = 1,2$ is the lattice number, and $(x,y)$ is the coordinate in the sample plane. The transmission function of the sample is given by the product of the two transmission functions corresponding to the two lattices

$$t(x,y) = t_1(x,y) t_2(x,y), \quad (2)$$

which can be expanded as:

$$t(x,y) = [1 - w_1(x,y)][1 - w_2(x,y)] = 1 - w_1(x,y) - w_2(x,y). \quad (3)$$

In Eq. 3 we took into account that $w_1(x,y)w_2(x,y) = 0$, this term is non-zero only when the two lattices are identical and precisely on top of each other, we do not consider such a situation because it does not create a moiré structure. The complex-valued far-field distribution of the scattered wave is given by the Fourier transform of the transmission function of the sample $t(x,y)$. The diffraction pattern is given by the square of the amplitude of the far-field wavefront distribution. From Eq. 3, it is apparent that the diffraction pattern will exhibit peaks due to diffraction on individual lattices but there will be no diffraction peaks due to formed moiré structure (moiré peaks).

Next, we assume that each particle has a finite size, so that its absorption can be described by a function $a(x,y)$. The transmission function of each lattice can be written as

$$t_i(x,y) = 1 - w_i(x,y) \otimes a(x,y), \tag{4}$$

and the transmission function of the sample is given by:

$$t(x,y) = [1 - w_1(x,y) \otimes a(x,y)][1 - w_2(x,y) \otimes a(x,y)] = \\ 1 - w_1(x,y) \otimes a(x,y) - w_2(x,y) \otimes a(x,y) + [w_1(x,y) \otimes a(x,y)][w_2(x,y) \otimes a(x,y)]. \tag{5}$$

Now the diffraction pattern will exhibit peaks due to diffraction on individual lattices and it will also exhibit peaks due to formed moiré structure (moiré peaks) as defined by the last term in Eq. 5.

The corresponding simulations are shown in Fig. 1. The simulations were done as follows. Two arrays of particles positions, corresponding to two identical lattices with a relative rotation of 10°, were created, giving $w_i(x,y)$. For each particle position, a spherical wave originating at this position was assumed. The far-field distribution of the wavefront scattered by lattice $i$ was calculated as $W_i(k_x, k_y) = \sum_n \exp\left[-i\left(k_x x_n^{(i)} + k_y y_n^{(i)}\right)\right]$, where $(x_n, y_n)$ is the coordinate of particle $n$. The inverse Fourier transform of the far-field distribution $W_i(k_x, k_y)$ gives $w_i(x,y)$. For the case, when the particles were assumed to have infinitesimal size, the transmission function of the lattice was calculated as $t_i(x,y) = 1 - w_i(x,y)$. For the case, when the particles were assumed to have finite size, $w_i(x,y) \otimes a(x,y)$ was calculated by applying the convolution theorem as $\text{FT}^{-1}\left[W_i(k_x, k_y) A(k_x, k_y)\right]$, where $A(k_x, k_y)$ was calculated as the Fourier transform of $a(x,y)$. The transmission function of the lattice was calculated as $t_i(x,y) = 1 - w_i(x,y) \otimes a(x,y)$. The transmission function of the sample $t(x,y)$ was calculated as the product $t_1(x,y) t_2(x,y)$. The diffraction pattern was calculated as the square of the absolute value of the Fourier transform of $t(x,y)$. For infinitesimal-size particles, the simulations are shown in Fig. 1a – c. The simulated diffraction pattern (Fig. 1c) exhibits only the peaks originating from the two lattices. For finite-size particles, the simulations are shown in Fig. 1d – f. The simulated diffraction pattern exhibits the peaks originating from the two lattices and the moiré peaks.

We therefore conclude that only finite size of a particle can explain the presence of moiré peaks in the diffraction pattern. For realistic lattices made up of atoms, the overlap of the atomic functions creates the periodical structure (moiré structure) which creates additional diffraction peaks (moiré peaks) in the diffraction pattern.

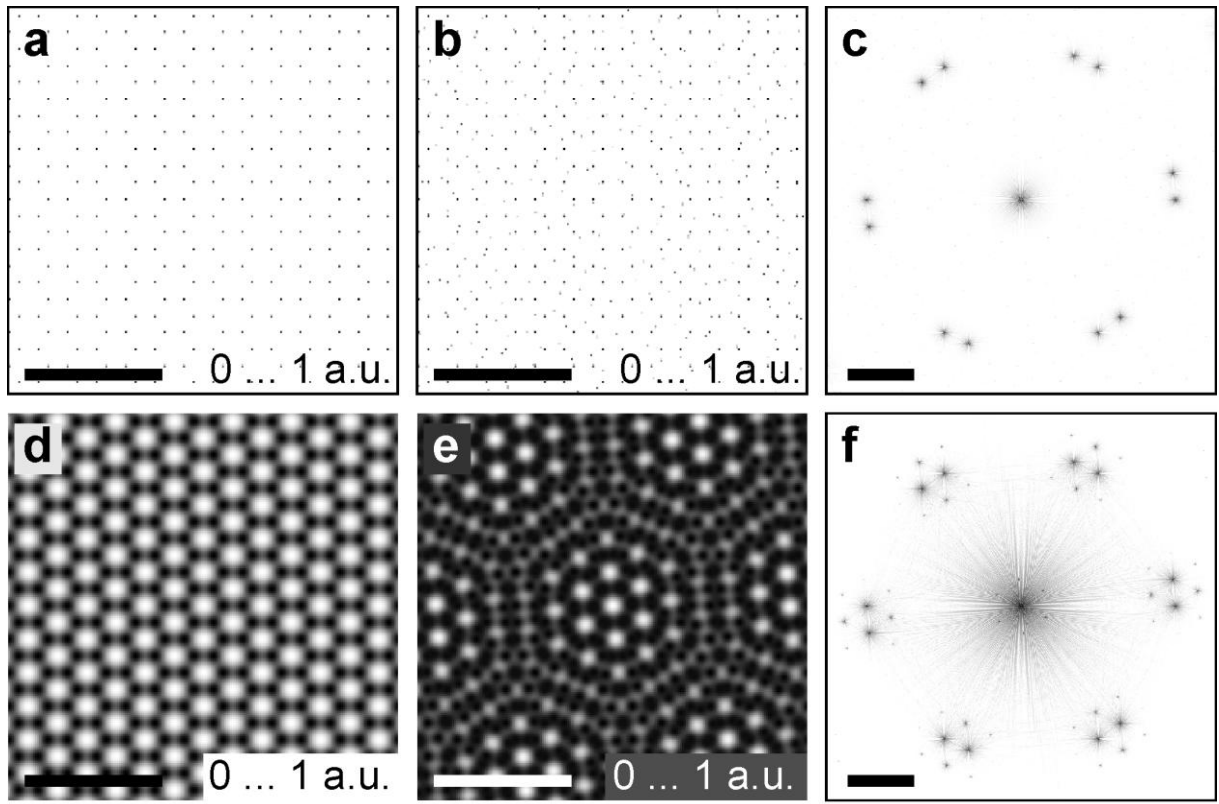

Fig. 1. Formation of moiré peaks in diffraction pattern. (a) Amplitude of the transmission function of hexagonal lattice made up of absorbing particles of infinitesimal size which can be represented as $\delta$-functions. (b) Amplitude of the transmission function of the sample consisting of two lattices shown in (a) with the twist angle of 10°. (c) Diffraction pattern of the sample shown in (b). (d) Transmission function of the hexagonal lattice made up of particles of finite size. Absorption function of each particle is described by a Gaussian distribution with the standard deviation of 0.5 Å. (e) Transmission function of the sample consisting of two lattices shown in (d) with the twist angle of 10°. (f) Diffraction pattern of the sample shown in (e). The particle arrangement and the inter-atomic distances in the individual lattices are the same as in graphene. The scalebar in (a), (b), (d) and (e) is 1 nm. The scalebar in (c) and (f) is 2 nm$^{-1}$.

## 2.2 Electron diffraction pattern of twisted bilayer graphene
### 2.2.1 Intensity of moiré peaks in diffraction pattern

In this section we consider realistic lattices, such as graphene lattices made up of carbon atoms. Electrons passing though a thin specimen only slightly deviate from their path. The effect of the specimen on the electron wave can be described by transmission function $t(x, y)$. A plane electron wave after passing through the specimen is described by:

$$t(x, y) = \exp[i\sigma V_z(x, y)] = \exp[i\sigma v_z(x, y) \otimes l(x, y)], \tag{6}$$

where $V_z(x, y)$ is the projected potential of the entire sample, $v_z(x, y)$ is the projected potential of an individual atom, $l(x, y)$ is the function describing positions of the atoms in the lattice, $\sigma = \dfrac{2\pi m e \lambda}{h^2}$ is the interaction parameter, $m$ is the relativistic mass of the electron, $e$ is the elementary charge, $\lambda$ is the wavelength of the electrons, $h$ is the Planck constant. The sample projected potential $V_z(x, y)$ does not depend on the energy of the probing electrons and it is only defined by the chemical origin of the atoms, while the interaction parameter $\sigma$ depends on the energy of the probing electrons. Simulated potential distribution $V_z(x, y)$ for graphene is shown in Fig. 2a, the details of the simulation are provided in Appendix A.

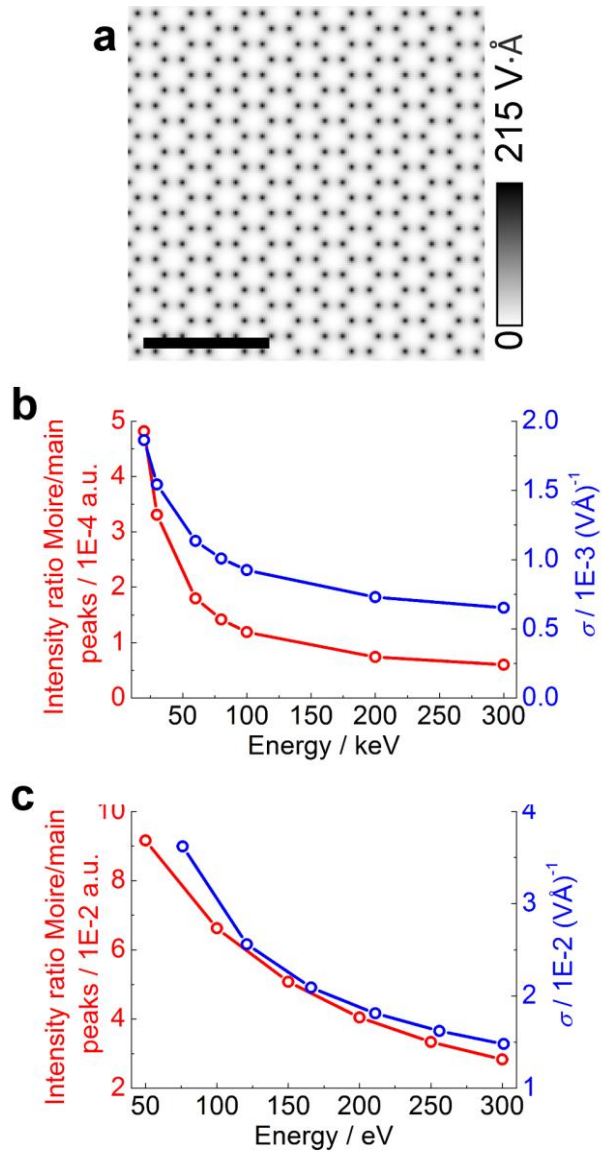

Fig. 2. Intensity of moiré peaks in the diffraction pattern. (a) Simulated projected potential distribution $V_z(x, y)$ for graphene. The scalebar is 1 nm. (b) and (c) averaged intensity ratio (moiré/first-order peaks), obtained from simulated diffraction patterns of TBG, and the interaction parameter $\sigma$ for high and low energy electrons as functions of the electron energy.

For TBG, for approximation of a weak phase object and neglecting the diffraction effects due to propagation between the two layers, the sample can be assigned the following transmission function:

$$\begin{aligned}
t(x,y) &\approx \left[1 - i\sigma v_z(x,y) \otimes l_1(x,y)\right]\left[1 - i\sigma v_z(x,y) \otimes l_2(x,y)\right] = \\
&\quad 1 - i\sigma v_z(x,y) \otimes l_1(x,y) - i\sigma v_z(x,y) \otimes l_2(x,y) - \\
&\quad \sigma^2 \left[v_z(x,y) \otimes l_1(x,y)\right]\left[v_z(x,y) \otimes l_2(x,y)\right].
\end{aligned} \quad (7)$$

The last term describes the overlap of the two lattices potentials which leads to the formation of the moiré structure and the corresponding moiré peaks in the diffraction pattern. From Eq. 7, it follows that the ratio between the main and the moiré peaks in the diffraction pattern (described by the second-, third and the last terms in Eq. 7, respectively) is approximately linearly proportional to the interaction parameter $\sigma$.

Diffraction patterns of TBG were simulated at different electron energies. In the simulation, the propagation between the layers was included, where the distance between the layers was set to 3.35 Å. The details of the simulation are provided in Appendix B. From the simulated diffraction patterns, the intensity ratio (moiré/first-order peaks) averaged over the six peaks was calculated and plotted as a function of electron energy, shown in Fig. 2b and c. The intensity ratio exhibits a similar dependency on the electron energy as the interaction parameter $\sigma$, as also described by Eq. 7. The interaction parameter $\sigma$ is relatively small for typical transmission electron microscope (TEM) electron energies (20 – 300 keV). For example, σ = 0.81 (keV·Å)$^{-1}$ for 100 keV. This can be the reason why moiré peaks have not been observed in the diffraction patterns of TBG acquired at typical TEM electron energies [11, 12]. At low electron energies, $\sigma$ is relatively high, for example, σ = 25.61 (keV·Å)$^{-1}$ at 100 eV and moiré peaks should be visible in the diffraction patterns. Below we show that moiré peaks indeed can be experimentally detected in diffraction patterns of TBG acquired with low-energy electrons.

## 2.3 Period and relative rotation of moiré structure

The period and the relative rotation of the moiré structure in a bilayer with the twist angle $\varphi$ (Fig. 3a) can be found by considering the wavevectors of the diffracted waves in reciprocal space, as shown in Fig. 3b. In the reciprocal space, the moiré structure wavevector is given by the difference between the two wave vectors corresponding to the two lattices: $\vec{\kappa} = \vec{k}_2 - \vec{k}_1$, its absolute value can be found from: $\kappa^2 = k_2^2 + k_1^2 - 2k_2 k_1 \cos\varphi$. By substituting the values of the wavevectors expressed through the periods as $k_1 = \frac{2\pi}{d_1}$, $k_1 = \frac{2\pi}{d_2}$, and $\kappa = \frac{2\pi}{d_M}$, we obtain the period of the moiré structure $d_M = \frac{d_2 d_1}{\sqrt{d_2^2 + d_1^2 - 2d_2 d_1 \cos\varphi}}$, or $d_M = \frac{d_1(1+\delta)}{\sqrt{2(1+\delta)(1-\cos\varphi)+\delta^2}}$, where we introduced the mismatch parameter $\delta$ as $d_2 = d_1(1+\delta)$. The moiré peaks are found at $\kappa = \frac{2\pi}{d_M}$, where $d_M$ is the moiré structure period, similar to the definition of the period $d$ in graphene, as illustrated in Fig. 3a. Instead of the periods of the structures, the same expression can be written for the lattices constants:

$$a_M = \frac{a_1(1+\delta)}{\sqrt{2(1+\delta)(1-\cos\varphi)+\delta^2}} = f_M a_1, \tag{8}$$

where we introduced the scaling factor $f_M$. The moiré structure rotation angle $\theta$ can be found using the sine rule $\frac{\kappa}{\sin\varphi} = \frac{k_2}{\sin\theta}$. $\theta$ is given by

$$\tan\theta = \frac{\sin\varphi}{(1+\delta)-\cos\varphi}, \tag{9}$$

and illustrated in Fig. 3a. Plots of the period and the angle of the relative rotation of the moiré pattern as functions of the twist angle $\varphi$ can be found in Ref. [13].

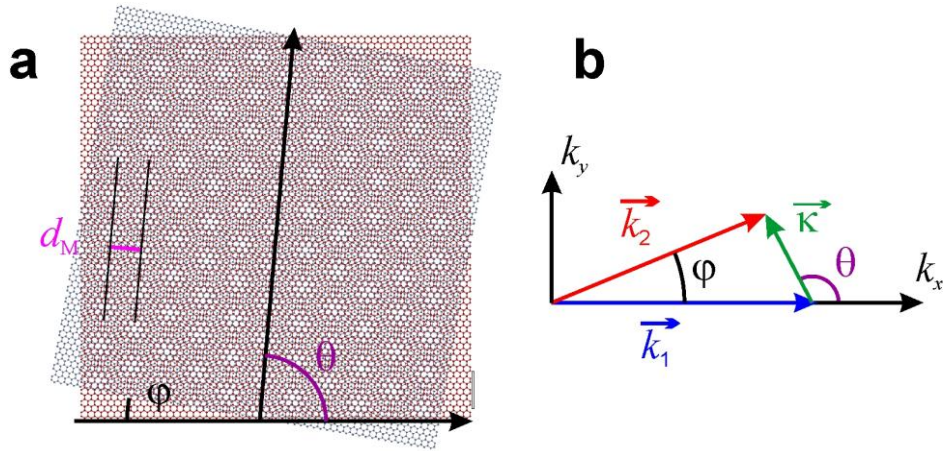

Fig. 3. Moiré pattern formation in a twisted bilayer. (a) Sketch of twisted bilayer, $\varphi$ is the twist angle. The period of the formed moiré structure $d_M$ and the relative rotation angle of the moiré structure $\theta$ are shown. (b) Vectors in the Fourier space, indicating the positions of the diffraction and the moiré peaks.

## 2.4 Low-energy electron experiments
### 2.4.1 Diffraction pattern of twisted bilayer graphene

The two low-energy electron microscopes employed in this study have been described in details in previous publications [14-16]. The source of the coherent electron beam is a sharp W(111) tip. The position of the tip was controlled by a 3-axis piezo-manipulator at nanometer precision. The tip was set at a negative voltage and the electrons were extracted by field emission [17]. In the diffraction mode, the electron beam was collimated by a microlens [18], as illustrated in Fig. 4a. The wave transmitted through the sample propagated to the detector unit where its intensity was recorded. The detector unit was positioned at 68 mm in the diffraction setup and consisted of a microchannel plate (MCP), a phosphor screen, and a digital camera. Graphene samples were prepared as described in Ref.[19].

Fig. 4b shows the experimentally acquired low-energy diffraction pattern of TBG. In this diffraction pattern, two sets of the intense first-order diffraction peaks with a relative rotation of about 17° are observed. Besides these intense first-order peaks, numerous weaker peaks are also observed. The positions of these weak peaks correspond to the positions of the moiré peaks associated with the twist angle $\varphi = 17°$, as confirmed by a simulation shown in Fig. 4c. The positions of the moiré peaks can be found by connecting the first-order peaks of the two lattices and translating the obtained vectors to the centre of the diffraction pattern (as indicated by the cyan and the green lines in Fig. 4c). The moiré peaks appear not only in the centre of the diffraction pattern, but the same set of moiré peaks appears centred at each first (or higher)-order diffraction peak. This leads to numerous weak intensity peaks observed in the diffraction pattern.

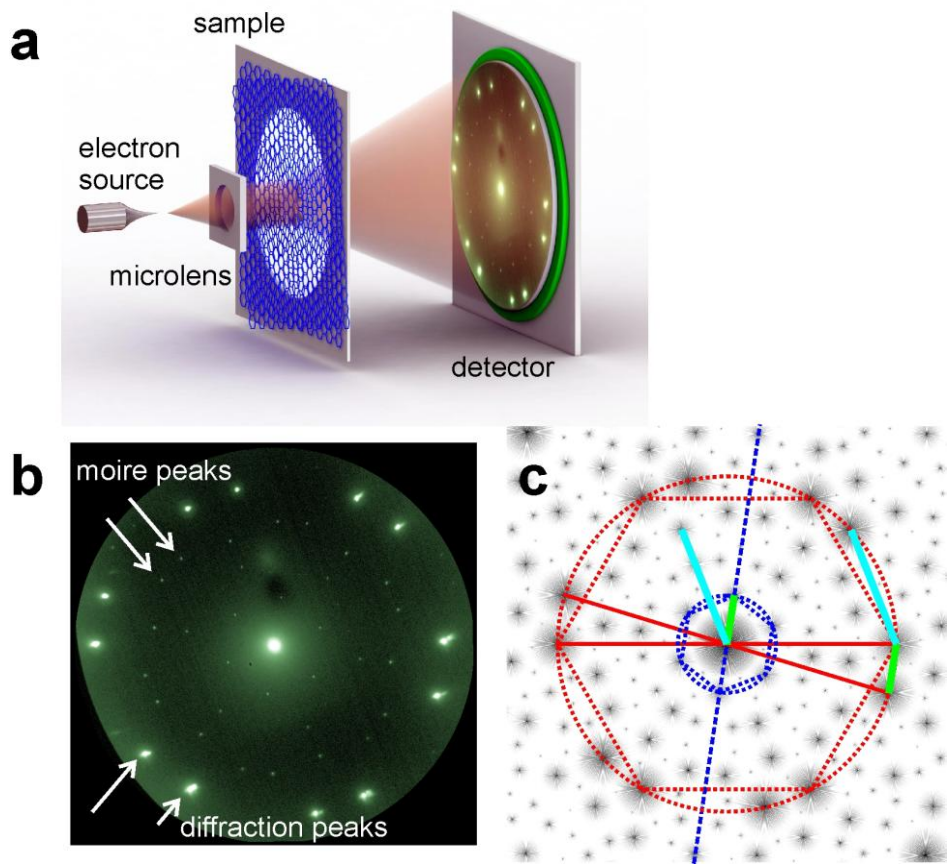

Fig. 4. Moiré pattern in bilayer graphene. (a) Sketch of the experimental setup for low-energy electron diffraction. (b) Experimental low-energy electron diffraction pattern of twisted bilayer graphene (TBG) acquired with 236 eV energy electrons, where moiré peaks are observed. Here the estimated twist angle is about $\varphi = 17°$. (c) Simulated diffraction pattern of graphene bilayer with the twist angle $\varphi = 17°$, the inverted intensity is shown in logarithmic scale. The cyan and the green lines indicate the distance between the first-order diffraction peaks which, when translated to the center of the diffraction pattern, provide the positions of the moiré peaks.

### 2.4.2 In-line hologram of twisted bilayer graphene

Images of TBG were also acquired in in-line holography mode, where a divergent electron wave is passing through the sample and its intensity is detected in the far field. The corresponding experimental setup is sketched in Fig. 5a. This experimental arrangement is similar to the original scheme of Gabor holography [20, 21], for this reason we call these images holograms. It should also be mentioned that imaging crystalline samples with a divergent/convergent electron wave constitutes the convergent beam electron diffraction (CBED) technique [22, 23]. CBED was successfully employed over the past few decades to investigate the structure of three-dimensional crystals [24-27]. The interpretation of CBED images of crystalline samples is not straightforward and typically consists of a simulation of the positions of higher-order Laue zone (HOLZ) lines by employing the Bragg law until the simulated positions match the HOLZ lines observed in the experimental CBED patterns [28]. The interpretation of CBED images of two-dimensional crystals is somewhat more straightforward and the reconstruction methods from holography can be adapted [29-31].

A low-energy electron hologram of a few layer twisted graphene (FLTG) sample is shown in Fig. 5b – c. The acquired hologram was normalized so that the average intensity of the region with monolayer graphene (MLG) amounts to 1. The intensity values are ranging from 0.34 to 0.58 a.u. An intensity profile through the intensity distribution (shown in Fig. 5d) exhibits a periodical dependency, which was fitted by a cosine function:

$$I(X) = I_0 \cos\left(\frac{2\pi}{T_X} X\right) + I_1, \tag{10}$$

with $I_0 = 0.12$ a.u., $T_X = 29$ nm and $I_1 = 0.46$ a.u.

It should be emphasized that in the in-line holography mode the intensity distribution is acquired in the far-field. Here the interpretation of the moiré pattern in *the far-field* is different from the interpretation of the moiré pattern obtained in the vicinity of the sample as it is for example done in STM [9, 10]. For example, if the atoms behave as only-phase objects, no change in the intensity of the transmitted wave would be detected right behind the sample. However, in the far field the intensity of the transmitted wave will exhibit variations that correspond to the moiré structure, as shown in Fig. 5b, c and e.

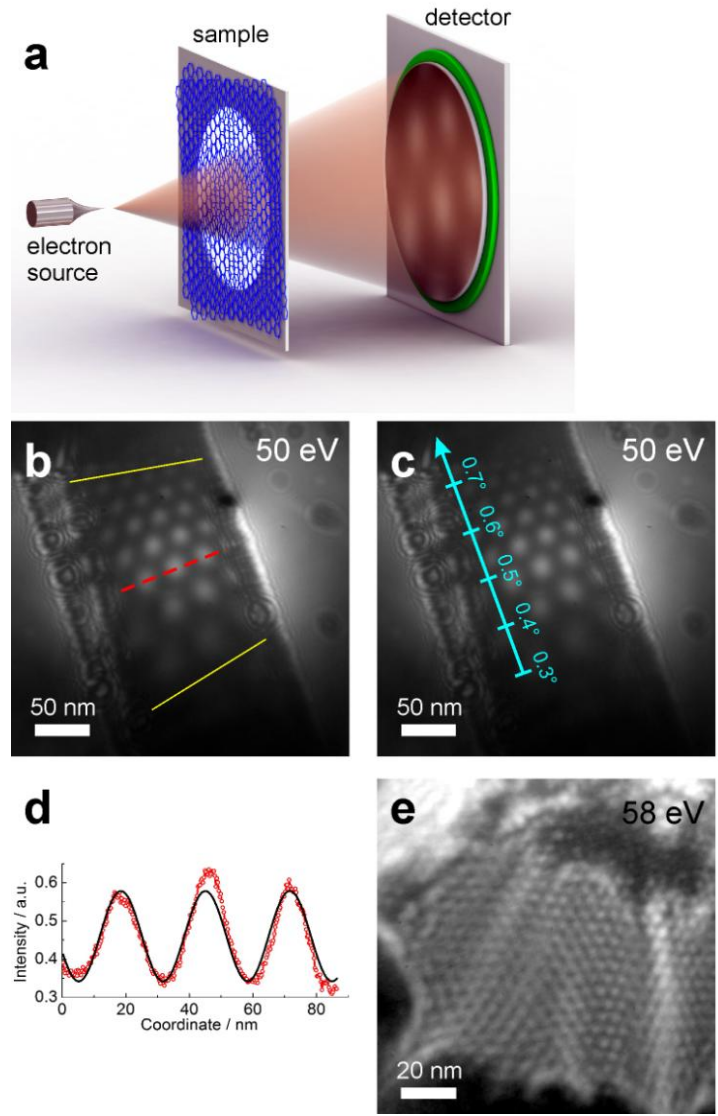

Fig. 5. Low-energy electron in-line hologram of few layer twisted graphene (FLTG). (a) Sketch of the experimental arrangement for in-line holography with low-energy electrons. (b) Low-energy electron in-line hologram of FLTG with two graphene layers creating a moiré structure. The hologram was acquired with 50 eV energy electrons, at the source-to-sample distance of 390 nm (the source-to-sample distance was obtained from the reconstruction). The yellow lines indicate the relative rotation of the moiré interference pattern. (c) The same as (b) with the indicated twist angle between the layers. (d) Intensity profile along the red dashed line in (b) and the fit obtained by Eq. 10. (e) Low-energy electron hologram of FLTG with two graphene layers creating a moiré structure with different contrast due to rippling. The hologram was acquired with

58 eV energy electrons, at the source-to-sample distance of 560 nm (the source-to-sample distance was obtained from the reconstruction).

# 3. THEORY AND CALCULATIONS

## 3.1 In-line holograms of twisted bilayer graphene

A moiré structure in TBG contains regions of different atomic arrangement. Atomic arrangement is approximately AA stacking in the moiré structure lobes, while in the regions between the moiré structure lobes the atoms are in approximately AB stacking, as illustrated in Fig. 6a.

Holograms of TBG were simulated assuming the kinematic first-order Born scattering approximation (the details of the simulations are provided in Appendix C). The simulated holograms at different interlayer distances $d$ are shown in Fig. 6b. The contrast in the holograms is changing with $d$, which can be explained by an additional phase shift that is a function of the interlayer distance $d$. An experimental hologram of the moiré structure where the contrast is changing as a function of $d$ due to rippling is shown in Fig. 5e. Also, in the simulated holograms of TBG, the resulting interference distribution exhibits some rotation and demagnification with increasing $d$ (Fig. 6b), which can be explained as follows. The electron wavefront after passing the first layer propagates towards the second layer. Since the in-line hologram is created with a divergent electron wave (as illustrated in Fig. 5a), in the plane of the second layer, the incident wavefront is modulated in such a way that it contains information about the first layer magnified by the factor $\frac{z_0 + d}{z_0} = 1 + \delta$, where $\delta = \frac{d}{z_0}$, and $z_0$ is the distance between the source and the first layer of the sample. $\delta$ can be as large as a few percent for typical source-to-sample distances of tens of nanometers, and therefore cannot be neglected. Fig. 6c and d show $f_M$ and $\theta$ as functions of the interlayer distance $d$. It is interesting to note that even when the bilayer consists of two identical non-twisted layers separated by a distance $d$, a moiré structure is formed. It is because the magnified projection image of the first layer lattice is superimposed onto the second layer lattice, and thus a moiré structure is formed due to the "lattice constant mismatch" $\delta = \frac{d}{z_0}$.

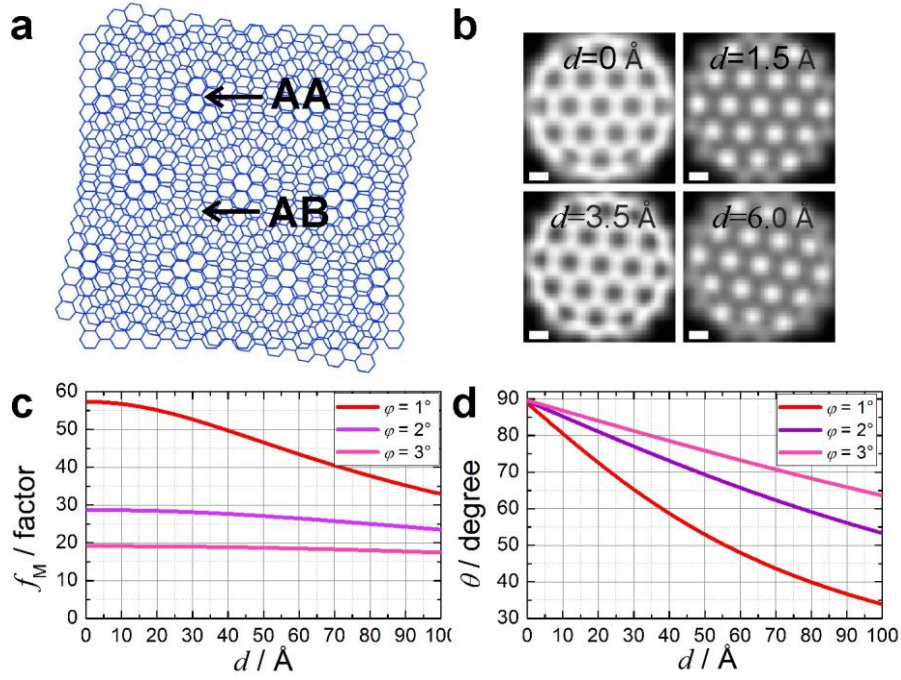

Fig. 6. Simulated in-line holograms of twisted bilayer graphene (TBG). (a) Illustration of the moiré structure in TBG, regions of AA and AB stacking are indicated. (b) Simulated intensity distributions in the far-field of TBG with various distances between the layers: $d = 0$, 1.5, 3.5 and 6 Å. The relative rotation between the layers is $\varphi = 1°$. The scalebars are 10 nm. (c) and (d) $f_M$ factor and the rotation angle $\theta$ as functions of the distance between the layers in TBG, calculated with Eq. 8 and 9, respectively. The simulations were done for 50 eV electrons, and the source-to-the first layer distance was set to 390 nm.

Next, we performed simulations to obtain the far-field intensity distributions similar to that observed in the experiment (Fig. 5b – c). The results are shown in Fig. 7. Since the experimental holograms exhibit only the change in the period and the relative rotation of the more structure, but not in the contrast, we assumed that the distance between the layers was the same. A matching contrast was obtained at $d = 6$ Å. Fig. 7a and b show $f_M$ and $\theta$ as functions of the twist angle $\varphi$. From the experimental hologram (Fig. 5b – c) we evaluate the change $f_M = 203$ to 81, while $\theta$ changes by about 20° (indicated by the yellow lines in Fig. 5b). These changes correspond to the change of the twist angle $\varphi$ approximately from 0.3° to 0.7°, as indicated in Fig. 5c. This range includes some of the "magic angles" (1.05°, 0.5°, 0.35°, 0.24° and 0.2°) at which the band velocity at the Dirac point turns into zero [7] and TBG demonstrates unusual electronic properties [6] such as super-conductivity [8]. Simulated holograms at a few different $\varphi$ are shown in Fig. 7c.

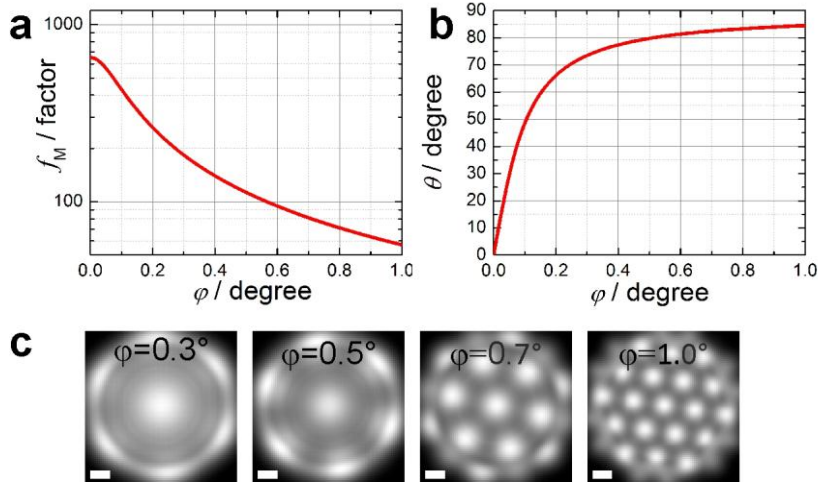

Fig. 7. Simulated in-line holograms of twisted bilayer graphene (TBG). (a) and (b) $f_M$ factor and the rotation angle $\theta$ as functions of the twist angle $\varphi$, calculated with Eq. 8 and 9, respectively. (c) Simulated far-field intensity distribution of TBG with various twist angle $\varphi = 0.3°$, 0.5°, 0.7° and 1.0°. The distance between the layers is $d = 6$ Å. The scalebars are 10 nm. The simulations were done for 50 eV electrons, and a source-to-the first layer distance of 390 nm.

We also performed additional simulations by adding an amplitude part to the transmission function of an individual atom in form of

$$a(x,y) = 1 - a_0 \exp\left(-\frac{x^2 + y^2}{2\sigma_a^2}\right), \tag{11}$$

where $a_0$ and $\sigma_a$ are the amplitude and the standard deviation, respectively. For $a_0 = 1$ and $\sigma_a = 0.2$ Å, we obtained almost the same intensity distributions as with amplitude $a(x,y) = 1$. We therefore conclude that the transmission function of an individual atom in form of a phase-only function provides a good match to the experimental observations.

## 4. DISCUSSION

We showed that the transmission function of an individual atom described by an only phase-shifting function matches well the experimentally observed intensity distributions in diffraction patterns and in-line holograms of moiré structures. It is somewhat surprising that a decrease of intensity in the far-field must not be associated with some absorption inside the sample but can simply be caused by a particular atomic arrangement, where each atom only slightly changes the phase of the electron wave. The total intensity is preserved and re-distributed between the zero and first or higher-order diffraction peaks or spots for diffraction and in-line holography modes respectively. The transmission properties of two or more layers graphene samples, such as absorption per layer [32], thus cannot be judged from the intensity distribution acquired in the far-field, since the latter is mainly defined by the arrangement of atoms in the layers. For example, BLG in AA and AB stacking will result in different intensities in the far-field. These findings can be useful when considering graphene as a sample support [33] in imaging experiments [34].

For typical high-energy electrons energies (80 – 300 keV) the interaction parameter $\sigma$ is so weak that the contrast of the far-field image of a moiré structure is too weak to be detected. Indeed, at typical TEM electron energies, no moiré peaks have been observed in the diffraction patterns of TBG [11, 12] and no contrast variations were observed in the zero-order CBED spots in the CBED patterns of TBG [30, 31]. However, for electrons of the intermediate energies, as for example at 30 keV, the weak phase approximation can be applied and, at the same time, sufficient contrast for detection of intensity variations due to the diffraction on moiré structure can be expected in possible future lower voltage TEM studies [35-40]. This would allow comparing the experimentally measured intensity values with the analytical solution, and the exact distributions of the transmission function of individual atoms can be in principle obtained.

# APPENDICES

APPENDIX A: Simulation of transmission function of graphene

The transmission function of a graphene layer was assumed in the form [41]:

$$t(x,y) = \exp[i\sigma V_z(x,y)] = \exp[i\sigma v_z(x,y) \otimes l(x,y)], \quad \text{(A1)}$$

where $V_z(x,y)$ is the projected potential of the entire sample, $v_z(x,y)$ is the projected potential of an individual atom, $l(x,y)$ is the function describing positions of the atoms in the lattice, $(x,y)$ is the coordinate in the sample plane, and $\otimes$ denotes convolution. No weak phase approximation was applied. The projected potential of a single carbon atom was simulated in the form [41]:

$$v_z(r) = 4\pi^2 a_0 e \sum_i a_i K_0\left(2\pi r \sqrt{b_i}\right) + 2\pi a_0 e \sum_i \frac{c_i}{d_i} \exp\left(-\pi^2 r^2 / d_i\right), \quad \text{(A2)}$$

where $r = \sqrt{x^2 + y^2}$, $a_0$ is the Bohr' radius, $e$ is the elementary charge, $K_0(...)$ is the modified Bessel function, and $a_i, b_i, c_i, d_i$ are the parameters tabulated in Ref. [41]. The analytical expression for $v_z(r)$ has singularity at $r = 0$, but because atom has a finite size with the radius of approximately $r = 0.1$ Å, $v_z(r)$ at $r = 0$ was replaced by the value at $r = 0.1$ Å, which is approximately 215 V·Å.

In Eq. A1, The convolution $v_z(x, y) \otimes l(x, y)$ was calculated by applying the convolution theorem as $\text{FT}^{-1}\{\text{FT}[v_z(x, y)]\text{FT}[l(x, y)]\}$, where FT denotes Fourier transform. $\text{FT}[l(x, y)]$ was simulated without applying Fast Fourier transforms (FFT) to avoid artifacts associated with FFT. $\text{FT}[l(x, y)]$ was simulated as $\text{FT}[l(x, y)] = \sum_n \exp[-i(k_x x_n + k_y y_n)]$, where $(x_n, y_n)$ are the atomic positions, $\Delta k$ was selected to be $1.25 \cdot 10^7$ m$^{-1}$, which relates to a sample size of 80 × 80 nm$^2$. The distributions were sampled with 5634 × 5634 pixels, which in the sample plane relates to 1 pixel = 0.142 × 0.142 Å$^2$. The inverse Fourier transform was calculated by applying inverse FFT to the product of $\text{FT}[v_z(x, y)]$ and $\text{FT}[l(x, y)]$.

## APPENDIX B: Simulation of diffraction patterns

The simulation of a diffraction pattern of TBG included the following steps. The two graphene layers in TBG were assigned transmission functions $t_1(x_1, y_1)$ and $t_2(x_2, y_2)$, respectively. The exit wave after passing through the first layer was given by $u_1(x_1, y_1) = t_1(x_1, y_1)$. Next, this wave was propagated to the second layer. The propagation was calculated by the angular spectrum method [41-43]. The propagated wave was described by the complex-valued distribution $u_2(x_2, y_2)$. The exit wave after passing through the second layer was calculated as $u_3(x_2, y_2) = u_2(x_2, y_2) t_2(x_2, y_2)$. The diffraction pattern was then simulated as the square of the amplitude of the Fourier transform of the exit wave $u_3(x_2, y_2)$.

## APPENDIX C: Simulation of in-line holograms

The simulation of an in-line hologram of TBG included the following steps. The two graphene layers in TBG were assigned transmission functions $t_1(x_1, y_1)$ and $t_2(x_2, y_2)$, respectively. The incident wavefront was simulated as

$$u_0(x_1, y_1) = \frac{1}{z_0} \exp\left[-\frac{x_1^2 + y_1^2}{2\sigma_0^2}\right] \exp\left[\frac{i\pi}{\lambda z_0}(x_1^2 + y_1^2)\right], \tag{C1}$$

which describes a spherical wave with a Gaussian-distributed amplitude, where $\sigma_0$ was selected such that the far-field intensity distribution would be matching the experimentally acquired intensity distribution, $\sigma_0 = 100$ nm at the source-to-sample distance of 390 nm. In addition, an apodization cosine window filter was applied to the resulting distribution, which gradually reduces the signal at the edges to zero and thus helps to minimize wrapping effect at the edges caused by the digital Fourier transform [43]. The exit wave after passing through the first layer was given by $u_1(x_1, y_1) = u_0(x, y) t_1(x_1, y_1)$. Next, this wave was propagated to the second layer. The propagation was calculated by the angular spectrum method [41-43]. The propagated wave was described by the complex-valued distribution $u_2(x_2, y_2)$. The exit wave after passing through the second layer was calculated as $u_3(x_2, y_2) = u_2(x_2, y_2) t_2(x_2, y_2)$. The hologram was then simulated as the square of the amplitude of the Fourier transform of $u_3(x_2, y_2)$, where the Fourier transform was calculated as FFT.

# ACKNOWLEDGEMENTS

Financial support of the University of Zurich is gratefully acknowledged. We are also grateful to Jean-Nicolas Longchamp and Flavio Wicki for acquiring the low-energy electron diffraction patterns and holograms.